\documentclass[10pt,a4paper]{article}
\usepackage[centertags]{amsmath}
\usepackage{amsfonts}
\usepackage{amssymb}
\usepackage{amsthm}
\usepackage{mathrsfs} %script fonts
\usepackage{epsfig,multicol}
\usepackage{graphicx}

\title{Quasi-normal modes, area spectra and multi-horizon spacetimes}

\author{Jozef Sk\'akala\footnote{jozef.skakala@ufabc.edu.br}\\
Centro de Matem\'atica, Computac\~ao e Cognic\~ao\\
Univesidade Federal do ABC\\
 Santo Andr\'e, S\~ao Paulo, Brazil}

\begin{document}

\maketitle

\begin{abstract}
We suggest an interpretation for the highly damped QNM frequencies of the spherically symmetric multi-horizon spacetimes (Reissner-Nord\-str\-\"om, Sch\-warzschild\--deSitter,\- Reissner-Nordstr\"om-deSitter) following Maggiore's proposal about the link between the asymptotic QNM frequencies and the black hole thermodynamics. We show that the behavior of the asymptotic frequencies is easy to understand if one assumes that all of the horizons have the same equispaced area spectra. The QNM analysis is then consistent with the choice of the area spectra to be the one originally proposed for the black hole's horizon by Bekenstein: $A=8\pi n$ ($l^{2}_{p}=1$). The interpretation of the highly damped QNM frequencies in the multi-horizon case is based on the similar grounds as in the single horizon (Schwarzschild) case, but it has some new features that are discussed in the paper.
\end{abstract}

\section{Introduction}
In the last more than a decade a considerable effort was concentrated on the question if there \emph{exists} any and \emph{what} is the role of the highly damped quasinormal modes (QNMs) in the area of black hole thermodynamics. The first conjecture about the link between the highly damped QNMs and the black hole thermodynamics was formulated by Hod \cite{Hod}. Hod's ideas caused significant interest in the problem within the scientific community and considerable amount of work followed Hod's original paper (e.g. see the papers \cite{Dreyer, Oppenheim, Zhang, Kiefer, Setare, Setare2, Hod2}). Later on Maggiore suggested interesting modification of Hod's conjecture \cite{Maggiore}. Maggiore's conjecture (or better Maggiore's modification of Hod's conjecture) finds the link between the asymptotic QNM frequencies and a quantum of energy emitted in a transition between black hole quantum states as:
\begin{equation}
\Delta M=\lim_{n\to\infty} \Delta_{(n,n-1)}~\omega_{nI}.
\end{equation}
(Everywhere in the paper we use the Planck units. Also one writes the complex QNM frequency as $\omega=\omega_{R}+i\cdot\omega_{I}$.)
In his original paper Maggiore applied his conjecture to derive the Schwarzschild black hole's horizon area spectrum as $8\pi n$ \cite{Maggiore} and more recently Maggiore's conjecture was applied for the same purpose in some other types of cases \cite{Vagenas2, Wei, Wei2, Lopez-Ortega1, Lopez-Ortega2}. 

For the spherically symmetric multi-horizon black hole spacetimes (in particular Reissner-Nordstr\"om, Schwarzschild-deSitter, Reissner-Nordstr\"om-deSitter black hole spacetime) the formulas for the highly damped QNM frequencies were derived in \cite{Motl, Andersson, Natario, Cardoso}. (Generally they are formulas for the highly damped QNM frequencies of the uncharged spin 0,1,2 perturbations.) The formulas for the asymptotic frequencies can be in all the three cases rewritten in the general form:
\begin{equation}\label{form}
\sum_{A=1}^{M}C_{A} \exp\left(\sum_{i=1}^{N}Z_{Ai}\frac{2\pi\omega}{|\kappa_{i}|}\right)=0.
\end{equation}
Here $Z_{Ai}$ are taking one of the values $Z_{Ai}=0,1,2$, furthermore $N$ is the number of horizons and $\kappa_{i}$ are the surface gravities of the horizons.

We analysed the behavior of the solutions of these formulas in \cite{Visser, Skakala1} with the following result: For all the cases (Reissner-Nordstr\"om, Schwarzschild-deSitter, Reissner-Nordstr\"om-deSitter) in case the ratios of all the surface gravities related to the different horizons are rational, the highly damped QNM frequencies can be split into finite number of equispaced families (labeled by $a$) of the form:
\begin{equation}\label{general}
\omega_{an}=(\hbox{offset})_{a}+i n \cdot\hbox {lcm}(|\kappa_{1}|, |\kappa_{2}|,...,|\kappa_{N}|).
\end{equation}
Here $\kappa_{1...N}$ are the surface gravities of the $N$ different horizons. By $lcm$ we mean the least common multiple of the numbers in the bracket\footnote{Note that in the general formula for the asymptotic frequencies derived in \cite{Visser} there is an additional parameter $g$ in the gap in the spacing between the frequencies. However analysis of the formulas obtained via monodromy calculations made in \cite{Skakala1} shows that in particular in the three relevant cases (R-N,S-dS,R-N-dS) the parameter $g$ has the value $g=2$ and one obtains the gap in the spacing as in \eqref{general}. This fact can be also easily observed from the formula \eqref{form}. 
Let us also add that in the reference \cite{Visser2} a qualitative analysis of the Schwarzschild-deSitter case has been done with a similar result to \eqref{form}. The only difference is that for $\kappa_{+}/\kappa_{-}=p_{+}/p_{-}$, ~$p_{+}\cdot p_{-}$~ even number, the $g$ parameter takes the value $g=1$ and this gives the spacing between the frequencies to be ~$2\cdot\hbox{ lcm}(\kappa_{+}, \kappa_{-})$~ instead of $\hbox{lcm}(\kappa_{+},\kappa_{-})$.}, hence ~
\[\hbox{lcm}(|\kappa_{1}|,|\kappa_{2}|,...,|\kappa_{N}|)=p_{1...N}|\kappa_{1...N}|,\]
where $p_{1...N}$ are relatively prime integers. If the surface gravity ratios are irrational there does \emph{not} exist an infinite periodic subset of QNM frequencies. (So the frequencies are in the strongest possible sense aperiodic.) One can see that the behavior of the frequencies in the multi-horizon case is considerably more complicated as in the single horizon case, but this is hardly anything surprising as the quantization of more than one horizon may play a role in the game.

The general behavior of the frequencies described by their splitting into subsets of the form \eqref{general} seems striking: It indicates that some important and interesting information might be encoded behind the formula \eqref{general}, on the other hand it seems to generally contradict Maggiore's conjecture. This contradiction can be exactly proven for the simplest case of scalar/electromagnetic-gravitational perturbations of the non-extremal Reissner-Nordstr\"om black hole \cite{Skakala2}. (Especially the case of Reissner-Nordstr\"om black hole is the one where one would \emph{not} expect the conjecture to fail as the thermodynamical variables are in this case well understood.) However in \cite{Skakala2} we left open possibilities how to try to reconcile Maggiore's conjecture with the behavior of QNM frequencies of the type \eqref{general}. Such reconciliation is the main goal of this work. So the purpose of this paper is to propose a way how to understand the general behavior of the frequencies \eqref{general} in the spirit of the Maggiore's conjecture, thus giving another evidence in favour of the conjecture. 

For already a long time it has been suggested that the black hole horizon's area should be quantized with an equidistant spectrum \cite{Bekenstein}. Let us write the general equidistant spectrum in a suggestive way:
\begin{equation}\label{area}
A=8\pi\gamma\cdot n, ~~~~~~~ n\in\mathbb{N}.
\end{equation}
 This type of area spectrum with $\gamma=1$ was obtained in many works (ie.\cite{Bekenstein, Barvinsky1, Barvinsky2, Medved, Vagenas}) and it is the one that was derived for Schwarzschild black hole via Maggiore's conjecture. (In fact it is the spectrum \eqref{area} with $\gamma=1$ that was derived via Maggiore's conjecture generically within the cases known to the author where the conjecture was successfully used.\footnote{The only exception known to the author is the paper \cite{Lopez-Ortega2} where the area quantum of the cosmological horizon in the deSitter spacetime is obtained as $16\pi=2\cdot 8\pi$ instead of $8\pi$. Whether the factor $2$ is a fundamental property of the cosmological horizon area spectrum, or only a result of the way the spacetime is in this case perturbed might be an open question.}) The main proposal of this paper leading to the understanding of formulas of the type \eqref{general} in the spirit of Maggiore's conjecture is that in the multi-horizon case the area spectra are of the form $8\pi n$ for each of the horizons in the multi-horizon spacetime.

\section{Non-extremal Reissner-Nordstr\"om black hole\label{sec2}} 
 Let us assume that the quantization \eqref{area} applies to all the horizons in the multi-horizon spacetime. For the non-extremal Reissner-Nordstr\"om black hole this means that both of the horizons have area given as $A_{\pm}=8\pi\gamma n_{\pm}$. (By $A_{+}$ we denote the area of the outer horizon, by $A_{-}$ the area of the inner Cauchy horizon.) In case of the presence of \emph{uncharged} electromagnetic-gravitational/scalar perturbations the only change of black hole parameters is given\footnote{We follow here the discussion in \cite{Natario}.} by $\Delta M$. Furthermore let us calculate the area difference that results from a quantum emission:
\begin{equation}\label{one}
\Delta A_{\pm}=8\pi r_{\pm}\frac{d r_{\pm}}{dM}\Delta M.
\end{equation}
$dr_{\pm}/dM$ can be expressed as
\begin{equation}\label{two}
\frac{dr_{\pm}}{dM}=-\frac{f'_{M}}{f'_{r}}|_{r_{\pm}}=\frac{2}{r_{\pm}f'_{r|r_{\pm}}}=\frac{1}{r_{\pm}\kappa_{\pm}}.
\end{equation}
(Here $f$ is the metric function for Reissner-Nordstr\"om spacetime given as $f(r)=1-2M/r+Q^{2}/r^{2}$.) The formulas \eqref{one} and \eqref{two} give the following result
\begin{equation}
\Delta A_{\pm}=8\pi \frac{\Delta M}{\kappa_{\pm}}.
\end{equation}
But now since both $\Delta A_{\pm}$ can be given only as $8\pi\gamma m_{\pm}$ ($m_{\pm}\in\mathbb{Z}$), then necessarily the following must hold:
\begin{equation}
\Delta M=\gamma \cdot m_{\pm}\kappa_{\pm}.
\end{equation}
This implies 
\begin{equation}
m_{+}\kappa_{+}=m_{-}\kappa_{-}~~~~~~\to~~~~~~\frac{\kappa_{+}}{\kappa_{-}}=\frac{m_{-}}{m_{+}},
\end{equation}
hence the surface gravities ratio must be rational. 

Let us stop here for a while: If both of the horizon's areas are quantized in the same way as \eqref{area}, then a transition, such that it changes only one black hole parameter $M$ leads to a change in the area of both of the horizons. Then one can obtain the area transitions consistent with the area spectra of both of the horizons only if the ratio of the two surface gravities is rational. Moreover the smallest allowed area transitions are no more necessarily given by $8\pi\gamma$ but by an integer multiples of this quantum, such that one can jump between different values of the spectra for both horizons in the same time.  
Now, one wants the quantum $\Delta M$ to be as small as possible, such that it does not violate the condition for the two horizon's area changes. This means:
\begin{equation}
\Delta M=\gamma\cdot\hbox{lcm}(|\kappa_{+}|,|\kappa_{-}|).
\end{equation}
But if Maggiore's conjecture is valid in this case, then necessarily the following must
hold
\begin{equation}
\lim_{n\to\infty}\Delta_{(n,n-1)}\omega_{nI}=\gamma\cdot\hbox{lcm}(|\kappa_{+}|,|\kappa_{-}|).
\end{equation}
Is this really the case? Not completely, but ``almost''. As already mentioned in the introduction it was shown that if the ratio of the two surface gravities of the two horizons is being rational, then asymptotic QNMs split into (generally) multiple families (marked by $a$) given as
\begin{equation}\label{families}
\omega_{an}=(\hbox{offset})_{a}+i n \cdot\hbox{lcm}(|\kappa_{+}|,|\kappa_{-}|).
\end{equation} 
Maggiore's conjecture in its original form suggests that one is supposed to monotonically order all the asymptotic QNM frequencies with respect to the imaginary part and then explore the $n\to\infty$ limit in the gap in the spacing in the imaginary part of the modes. This leads (at least for rational ratios $\kappa_{+}/\kappa_{-}=p_{+}/p_{-}$ with $p_{\pm}$ relatively prime and $p_{+}\cdot p_{-}$ odd) to a periodically changing gap and the limit in $\Delta_{(n,n-1)} \omega_{nI}$ does not exist \cite{Skakala2}. The splitting of the frequencies into \eqref{families} suggests that instead of $\Delta_{(n,n-1)}\omega_{nI}$ one has to rather consider $\Delta_{(n,n-1)}\omega_{anI}$, hence the gap in the spacing within each of the equispaced families. This means that transferring Maggiore's conjecture from the Schwarzschild case to the multi-horizon case in a straightforward way is misleading: only the equispaced subsets of QNM frequencies represent in the $n\to\infty$ limit the physical energies of the quanta emitted by the black hole. (On the other hand each frequency in the asymptotic sequence belongs to one of such subsets.)

Eventually, by extending Maggiore's conjecture to the multi-horizon case in the suggested way we use the equation \eqref{families} to obtain $\gamma=1$. This gives the horizon's area spectra to be exactly as expected:  $A=8\pi n$. (So it is fully consistent with the Schwarzschild spectrum obtained via Maggiore's conjecture. In the same time it gives the most ``popular'' option in the current literature.) 

Note the importance of the surface gravities rational ratios in the preceding argumentation: If the rational ratio condition would \emph{not} be fulfilled, then there is aperiodicity in an arbitrary infinite subsequence of QNM frequencies and moreover, the $n\to\infty$ limit of $\Delta_{(n,n-1)}\omega_{nI}$ does \emph{not} exist \cite{Skakala2}. (One can exactly prove the non-existence of such limit in the case of irrational ratios for the non-extremal Reissner-Nordstr\"om black hole, but it is very natural to expect that it holds also in the other spherically symmetric multi-horizon cases.) In such case there does not seem to be even a slightly modified way how to understand the highly damped modes via Maggiore's conjecture. The fact that the logic developed in this section guarantees the existence of the surface gravities rational ratios seems to be another coincidence that ``miraculously'' fits into the picture. (Let us mention that indications in favour of surface gravities rational ratios come also directly from the multi-horizon spacetime thermodynamics \cite{Padmanabhan}.) 

Thus let us conclude the following: Let us assume that 
\begin{itemize}
\item  each of the horizon's has quantized area as $8\pi n$, ($n\in\mathbb{N}$),
which implies rational ratios of surface gravities\footnote{At least if single black hole energy transitions are allowed.}, 
\item  in general the information about the mass quantum is contained only in subsequences of the infinite tower of the QNM frequencies. (But the union of the subsequences gives the full asymptotic sequence of the QNM frequencies.)  
\end{itemize}
Then under all these assumptions the behavior of the highly damped quasi-normal frequencies is fully explained.
Thus such ``fine'' modification of Maggiore's conjecture removes the trouble with asymptotic QNM frequencies of the electromagnetic-gravitational/scalar perturbations that was described in detail in \cite{Skakala2}.

\section{Other spherically symmetric multi-horizon spacetimes}
Moreover the power of our analysis from the section \ref{sec2} largely exceeds the non-extremal Reissner-Nordstr\"om case. Let us assume that we have radially symmetric black hole spacetime with the line element
\begin{equation}
-f(M,c_{i},r)~dt^{2}+f(M,c_{i},r)^ {-1}~dr^{2}+r^{2}d\Omega^{2}.
\end{equation}
By $c_{i}$ we mean some other parameters of the metric function, such as cosmological constant, charge (etc.). Assume that the function $f$ is given as
\begin{equation}\label{metriccondition}
f(M,c_{i},r)=-\frac{2M}{r}+\tilde f(c_{i},r).
\end{equation} 
(This is for example the case of Schwarzschild-deSitter spacetime and Reissner-Nordstr\"om-deSitter spacetime.) Then one can repeat all the previous calculations from the section \ref{sec2} and derive that if every horizon has the same equispaced area spectrum, then only the rational ratios of surface gravities are allowed. (Again one assumes that the underlying physics allows single $M$ parameter transitions.) Moreover, with our small modification / generalization of Maggiore's conjecture the QNM frequencies should split into families given as
\begin{equation}
\omega_{an}=(\hbox{offset})_{a}+i n\gamma\cdot\hbox {lcm}(|\kappa_{1}|, |\kappa_{2}|,...,|\kappa_{N}|),
\end{equation} 
with $\kappa_{1...N}$ being the surface gravities of the horizons of the multi-horizon spacetime. This is the case at least for the Schwarzschild-deSitter and Reissner-Nordstr\"om-deSitter spacetime. (The formula \eqref{general} holds for these two cases.) Moreover the formula \eqref{general} gives the value $\gamma=1$ for the area quantization of any horizon in any of these black hole spacetimes, thus the area spectrum of any of the horizons is $A=8\pi n$. 

\section{Conclusions}
In this paper we suggested an interpretation for the results for highly damped QNM frequencies of various spherically symmetric multi-horizon black hole spacetimes via the spirit of Maggiore's conjecture. (In particular we talk about Reissner-Nordstr\"om, Schwarzschild-deSitter and Reissner-\-Nordstr\"om-\-deSit\-ter\- spacetime.) We started with an assumption that all the horizons have the same equispaced area spectra. By using such an assumption we showed that there are allowed only rational ratios of the surface gravities in case the asymptotic quasinormal modes of some perturbing field relate to a change of only one of the black hole parameters. (Such a quantization of surface gravities / horizon temperatures on its own seems to be fascinating.) Then we could prove that (in the sense of slightly modified/generalized Maggiore's conjecture) the asymptotic frequencies of the spin 0,1,2 perturbations match with the horizon's area spectra given as $8\pi n$. (This was shown first for the non-extremal Reissner-Nordstr\"om spacetime where the thermodynamical interpretation is straightforward and then also for the asymptotically non-flat Schwarzschild-deSitter and Reissner-Nordstr\"om-deSitter spacetime.) One can expect that such a horizon area spectra will be a general consequence of the Maggiore's conjecture (at least) in case of spherically symmetric multi-horizon spacetimes. Thus also in the multi-horizon cases the (extended) Maggiore's conjecture provides information about the horizon's area spectra, only the way in which the information is encoded is (for quite obvious reasons) much more tricky as in the Schwarzschild single horizon case. 

\bigskip

{\bf Acknowledgments:} This research was supported by Fundac\~ao de Amparo a Pesquisa do Estado de S\~ao Paulo.

\end{document}